\documentclass[11pt, oneside]{article}
\pdfoutput=1

\usepackage{jheppub}
\usepackage{amsmath, amsthm, amssymb, graphicx, subfigure}

\newcommand{\Tr}{\operatorname{Tr}}

\preprint{DIAS-STP-18-09, KEK-TH-2053}

\author[a,b]{Yuhma Asano,}
\author[c]{Veselin G. Filev,}
\author[a]{Samuel Kov\'a\v{c}ik,}
\author[a]{Denjoe O'Connor,}
\affiliation[a]{School of Theoretical Physics,\\ 
       Dublin Institute for Advanced Studies, \\
       10 Burlington Road, 
       Dublin 4, Ireland.}
\affiliation[b]{KEK Theory Center, High Energy Accelerator Research Organization,\\
1-1 Oho, Tsukuba, Ibaraki 305-0801, Japan.}
\affiliation[c]{
Institute of Mathematics and Informatics,\\
Bulgarian Academy of Sciences,\\ Acad. G. Bonchev Str.,\\
1113 Sofia, Bulgaria.}
\emailAdd{yuhma@stp.dias.ie}
\emailAdd{vfilev@stp.dias.ie}
\emailAdd{skovacik@stp.dias.ie}
\emailAdd{denjoe@stp.dias.ie}

\abstract{We study the maximally supersymmetric plane wave matrix
  model (the BMN model) at finite temperature, $T$, and locate the
  high temperature phase boundary in the $(\mu,T)$ plane, where $\mu$
  is the mass parameter.  We find the first transition, as the system
  is cooled from high temperatures, is from an approximately $SO(9)$
  symmetric phase to one where three matrices expand to form fuzzy
  spheres. For $\mu > 3.0$ there is a second distinct transition at a
  lower temperature. The two transitions approach one another at
  smaller $\mu$ and merge in the vicinity of $\mu=3.0$. The resulting
  single transition curve then approaches the gauge/gravity prediction
  as $\mu$ is further decreased. We find a rough estimate of the
  transition, for all $\mu$, is given by a Pad\'e resummation of the
  large-$\mu$, 3-loop perturbative, predictions. We find evidence that
  the transition at small $\mu$ is to an M5-brane phase of the theory.}

\title{The non-perturbative phase diagram of the BMN matrix model}

\begin{document}

\maketitle

\section{Introduction}
The BMN or plane wave matrix model model \cite{Berenstein:2003gb} is
the matrix model description of M-theory in discrete lightcone quantisation
(DLCQ) when the M2-brane is propagating on an eleven dimensional
plane wave background. This background preserves the maximal,
sixteen, supersymmetries and is a massive deformation of the
BFSS model \cite{Banks:1996vh,Townsend:1995kk,Susskind:1997cw}.

In contrast to the BFSS model, the BMN matrix model has a discrete
energy spectrum. It is conjectured to capture the entire dynamics of
M-theory on the plane wave background and provide a non-perturbative
definition of M-theory itself.  Alternatively, it describes a system
of D0-branes of IIA supergravity. It has BPS ground
states\footnote{
The vacua of the BMN model correspond to $\frac{1}{2}$-BPS states 
of the eleven dimensional or IIA supergravity. 
They are invariant under infinitesimal transformations of the $SU(2|4)$ supergroup,
which has sixteen supercharges.
}, where the D0-branes expand to fuzzy spheres
\cite{Dasgupta:2002ru,Dasgupta:2002hx}.  In \cite{Maldacena:2002rb} it
is argued that the model also provides a regularisation of
NS5/M5-branes.

For large $\mu$ the model becomes a supersymmetric gauged Gaussian
model. Perturbation theory around this limiting model was performed up
to three loop order \cite{Spradlin:2004sx,Hadizadeh:2004bf} and a
large $\mu$ series for the transition in the Polyakov loop was
developed.  The dual gravitational theory was studied by perturbing
around the BFSS dual geometry to linear order in small $\mu$ in
\cite{Costa:2014wya} and the corresponding dual gravity prediction for
the transition was obtained to this order.

The gravity dual has a remarkably rich structure.  At zero temperature
it is described by bubbling geometries.  The dual bubbling geometries
are in eleven-dimensional supergravity compactified in one
translationally invariant direction on the two-dimensional subspace
\cite{LLM}, or equivalently, in type IIA supergravity \cite{LM}.
The supergravity solutions are labeled by configuration of
D2/M2 and NS5/M5 charges, discretely located on a one-dimensional
subspace of the geometries, and each of them corresponds to a vacuum
of the BMN model.  In the large $N$ limit there are infinitely many
solutions, and interesting special solutions, such as those
corresponding to a single stack of D2/M2-branes or NS5/M5-branes,
exist.  At finite temperature, the system has a Hawking-Page
transition---a transition between thermal and black hole spacetimes.
The solution examined in \cite{Costa:2014wya} has $S^8\times S^1$
horizon topology, which corresponds to the trivial vacuum in the BMN
model, where the $S^1$ is the M-theory circle.  It is surprisingly
difficult to obtain general gravity solutions dual to the BMN model
at finite temperature; even the solution in \cite{Costa:2014wya}
required numerical computations to obtain the free energy.

There is some understanding of the dual geometry from
the gauge-theory side.  Since the corresponding supergravity solutions are
supposed to be obtained at low energy, the geometry should be
constructed by a low-energy moduli operator in the gauge theory.  In
\cite{Asano:2014vba,Asano:2014eca}, the BPS operator that is expected
to pick up the low-energy moduli was computed by the supersymmetric
localisation method, and in the appropriate large-$N$ and strong
coupling limit, it was found that its eigenvalue distribution
satisfies the same integral equations as those
determining the supergravity solution. These equations govern a
non-trivial part of the dual metric, which is not determined by
the isometry\footnote{ Related with the isometry part, the spherical shell
  distributions of $S^2$ and $S^5$ can be obtained in the BMN model in
  the limits where M2- and M5-branes are realised, respectively.  
  The radii of these spheres completely agree with the M2 and M5 radii
  in the brane picture 
  \cite{Asano:2017xiy,Asano:2017nxw}.}.  Although these results
successfully reproduced part of the supergravity solution, the dynamics of
the emergence of those geometries, such as how the geometries are
favoured or superposed as the temperature changes, is beyond the scope
of such methods; therefore numerical simulations for the thermal
theory provide important information on which geometry emerges
naturally.

This paper is dedicated to obtaining a non-perturbative estimate of
the high temperature phase boundary as a function of the mass
parameter, $\mu$.

We find that a Pad\'e resummation of the large $\mu$
perturbative series \cite{Hadizadeh:2004bf}, with the assumption
that the transition temperature goes to zero linearly with $\mu$,
gives a result very close to that of the gravity calculations
\cite{Costa:2014wya} when expanded to linear order in small $\mu$. We
use this interpolating Pad\'e resummed curve as a guide to the
location of the transition.  In practice, we find it is an excellent
guide, giving a reasonable approximate location of the transition for
all values of $\mu$.

We study the BMN model non-perturbatively using Rational Hybrid Monte
Carlo techniques\footnote{A preliminary study of the BMN model concerning 
  a phase transition was carried out in \cite{Catterall:2010gf}. 
  The results differ significantly from ours at lower temperatures 
  due to, we believe, the larger lattice effects inherent in their first order code.
  In \cite{Kawahara:2006hs} and \cite{Ishiki:2008te,Ishiki:2009sg}, 
  the transition temperature was estimated by using an effective theory.
  Another numerical simulation of the full BMN model was done 
  around a special vacuum in \cite{Honda:2013nfa}.
    }.
We map the phase diagram in the $(\mu,T)$ plane and determine the phase
boundary as the system cools from the high temperature phase. We find
that for large $\mu$, a sharp transition in the Polyakov loop locates
the transition.  However, for $\mu\sim 7.0$ we find that there are two
phase boundaries and a new phase appears characterised by the Myers
cubic term in the action acquiring a significant non-zero vacuum
expectation value and the approximate $SO(9)$ symmetry no longer
holds.  We observe that this transition, which we call a Myers
transition, is from a phase where the system fluctuates around the
trivial configuration to one with fluctuations around expanded
fuzzy spheres. In the transition region we observe that the system also
fluctuates between different intermediate fuzzy sphere configurations.  The
transition should be in the universality class of the emergent
geometry transitions discussed in
\cite{DelgadilloBlando:2007vx,DelgadilloBlando:2008vi,OConnor:2013gfj}.
As the system is further cooled, with fixed $\mu$ (e.g.~see $\mu=6.0$
in Figure~\ref{Myers_Fig_N8μ6p0L24:Energy:Polyakov:R2ii}), the
Polyakov loop undergoes a separate transition but at a significantly
lower temperature.

For $\mu < 3.0$ we find the two transitions merge, and the
combined transition is visible in both the Myers observable and the
Polyakov loop and has a significant jump in the
energy, $E=\langle H\rangle/N^2$, where $H$ is the
Hamiltonian (see $\mu=2.0$ in Figure~\ref{Energy_Fig_N8μ2p0L24}).
For $\mu < 3.0$, the transition in the Polyakov loop is
rather difficult to observe but one can check
that the eigenvalue density of the holonomy undergoes a transition
from a gapped to a gapless spectrum as the system is cooled through the
Myers transition temperature.  

For small $\mu$, in the transition to a stable fuzzy sphere phase the
Polyakov loop tends to increase sharply (see Figure
\ref{EnergyAndPol_N6_μ2p0L24}) to a higher value before decreasing
slowly. In turn, for
fuzzy sphere backgrounds, the transition in the holonomy to a gapless phase
occurs at a considerably lower temperature than the Myers transition;
the latter transition coincides with the
region where the holonomy associated with the trivial background 
becomes gapless. 
These
different temperatures would correspond to different Hawking-Page
transitions for different solutions in the gravity dual.
This result is consistent with the upper bound for
the critical temperature discussed in \cite{Costa:2014wya}.

At large $N$ the system can be trapped in a given vacuum phase and
the transition between such vacua, being a quantum effect, is
suppressed.  This stability was used in \cite{Honda:2013nfa} to study
${\cal N}=4$ super Yang-Mills from the BMN model.  The parameter
region they used is located at lower temperatures than the Myers
transition we observe, and thus consistent with our findings.

A key ingredient in ensuring that the system was in the preferred
thermodynamic state, and not trapped in a false vacuum, was to cool the
system adiabatically so that $E(T)$ remained a monotonic function of
temperature.

In the intermediate $\mu$ regime the transition in the Myers term
occurs at a significantly higher temperature than that predicted by
our Pad\'e curve while the transition in the Polyakov loop tracks it
more closely.  For smaller $\mu<2$ the Myers transition again
approaches the Pad\'e curve and begins to qualitatively agree with the
dual gravity prediction \cite{Costa:2014wya}.  The observed
transition also is from small to large Myers term.  We
interpret this as meaning that on the gravity side a spherical
black hole becomes unstable to asymmetric deformations. To our knowledge,
the relevant gravity solution describing the black hole corresponding to a fuzzy sphere vacuum
has not yet been constructed.

The principal results of this paper are:
\begin{itemize}
\item{} A determination of the high temperature phase boundary
  of the BMN matrix model (see Figure~\ref{mu-T_Phase_diagram}).
\item{} A Pad\'e approximant estimate for the phase transition in the
  trivial vacuum.
\item{} A detailed non-perturbative study of the model for $N=8$.
\item{} An extrapolation to large $N$ of the Myers transition for $\mu=2.0$.
\item{} For $\mu=2.0$ we find the Myers observable takes a finite, $N$
  independent, non-zero value in the range studied. This suggests the
  transition can be interpreted as one to an M5-brane phase of the theory.
\end{itemize}

The paper is organised as follows. Section \ref{TheBMN_Model}
describes the model and the Pad\'e resummation of the large $\mu$
transition curve. Section \ref{LatticeFromulation} summarises our
lattice formulation of the model. Section \ref{Observables} discusses
the observables we measure. Section \ref{ResultsAndPhaseDiagram} gives
our results and describes the phase diagram we find. We end with some
concluding remarks and discussion.

\section{The BMN Model}
\label{TheBMN_Model}
The Euclidean thermal action for the BMN matrix model is given by 
\begin{align}
 S[X,\psi]
 =N\int_0^{\beta} d \tau  \, 
 \Tr \Bigg[ &
 \frac{1}{2}D_\tau X^i D_\tau X^i 
 -\frac{1}{4} \left( [X^r,X^s]+\frac{i\mu}{3}\varepsilon^{rst}X_t \right)^2
 \nonumber \\
 &
 -\frac{1}{2} [X^r,X^m]^2
 -\frac{1}{4} [X^m,X^n]^2
 +\frac{1}{2}\left( \frac{\mu}{6}\right)^2X_m^2
 \nonumber \\
 &
 +\frac{1}{2} \psi^T\mathcal{C}\left( D_\tau -\frac{i\mu}{4}\gamma^{567} \right) \psi
 -\frac{1}{2} \psi^T\mathcal{C}\gamma^i [X^i , \psi] 
 \Bigg] ,
 \label{BMNaction}
\end{align}
where $i,j=1,\cdots,9$, $r,s=5,6,7$ and $m,n=1,\cdots,4,8,9$. 
The covariant derivative is defined by $D_\tau\cdot=\partial_\tau\cdot-i[A,\cdot]$ and
$\mathcal{C}$ is the charge
conjugation matrix of $Spin(9)$. In (\ref{BMNaction}) we have
rescaled $X^i$, $\psi$, $\tau$ and $A$ to absorb the dependence
on the 't Hooft coupling defined as $\lambda=N g^2$ and we use
$\beta=\lambda^{1/3}/T$ and $\mu=\mu_0/\lambda^{1/3}$ with $\mu_0$ the
mass parameter of the plane wave geometry.

In the large $\mu$ limit the model (\ref{BMNaction}) reduces to a
supersymmetric Gaussian model. This simple Gaussian model 
undergoes a confinement-deconfinement phase transition
as the temperature is lowered and a straightforward calculation gives
the critical temperature in this limit as $T_c=\frac{\mu}{12\ln3}$.
This transition has been studied perturbatively in $1/\mu$
\cite{Spradlin:2004sx,Hadizadeh:2004bf}, where in a three loop calculation it was found that 
\begin{equation}
 T_c=\frac{\mu}{12\ln3}\left\{1+\frac{2^6\times 5}{3}\frac{1}{\mu^{3}}
  -\left(\frac{23\times19927}{2^2\times 3}+\frac{1765769\ln3}{2^4\times3^2}\right)\frac{1}{\mu^{6}}+\cdots\right\} .
 \label{LargemuTc} 
\end{equation}
This result, while reliable for large $\mu$ becomes untrustworthy as
$\mu$ decreases and passes through zero for $\mu\simeq13.4$. However,
if we perform a Pad\'e resummation of (\ref{LargemuTc}), assuming that
$T_c\rightarrow 0$ linearly with $\mu$ as $\mu\rightarrow0$, then we
can rewrite (\ref{LargemuTc}) as
\begin{equation}
  T_c=\frac{\mu}{12\ln3}\left\{1+\frac{\frac{320}{3\mu^{3}}}{1+\frac{3}{320\mu^{3}}(\frac{458321}{12}+\frac{1765769\ln 3}{144})+\cdots}
  \right\} ,
  \label{PadeResummedTc}
\end{equation}
which to linear order in small $\mu$ leads to the prediction
\begin{equation*}
 T_c=\frac{\mu}{12\ln3}\left(1+\frac{1638400}{5499852+1765769\ln 3}\right)\simeq 0.0925579 \mu .
\end{equation*}
This value is surprisingly close to
\begin{equation}
 T_c=0.105905(57)\mu ,
  \label{gaugegravity}
\end{equation}
which was obtained from a rather involved dual gravity computation \cite{Costa:2014wya}.

We use the Pad\'e resummed result (\ref{PadeResummedTc}) as a guide to
where one might expect the transition in the full model as $\mu$ is
decreased and study the system using the rational hybrid Monte Carlo
algorithm and a novel lattice discretisation described in
\cite{LatticePaper}. 

The action (\ref{BMNaction}) is minimised by the fuzzy sphere configurations,
\begin{equation}
  X^r=-\frac{\mu}{3} J^r \quad\hbox{and}\quad X^m=0 ,
  \label{FuzzySphereVac}
\end{equation}
where $J^r$ are generators of $SU(2)$ in an arbitrary representation of total
dimension $N$. These are BPS states and are protected ground states of the
quantum Hamiltonian.

\section{Lattice Formulation}
\label{LatticeFromulation}
We use a second order lattice discretisation of the model.  In this
formulation the time interval is replaced by a periodic lattice with
$\Lambda$ sites. The matrices are located on the lattice site, and the
lattice spacing is $a=\frac{\beta}{\Lambda}$.

\begin{figure}[t] 
  \begin{tabular}{@{}c@{}}
    \includegraphics[width=2.74in]{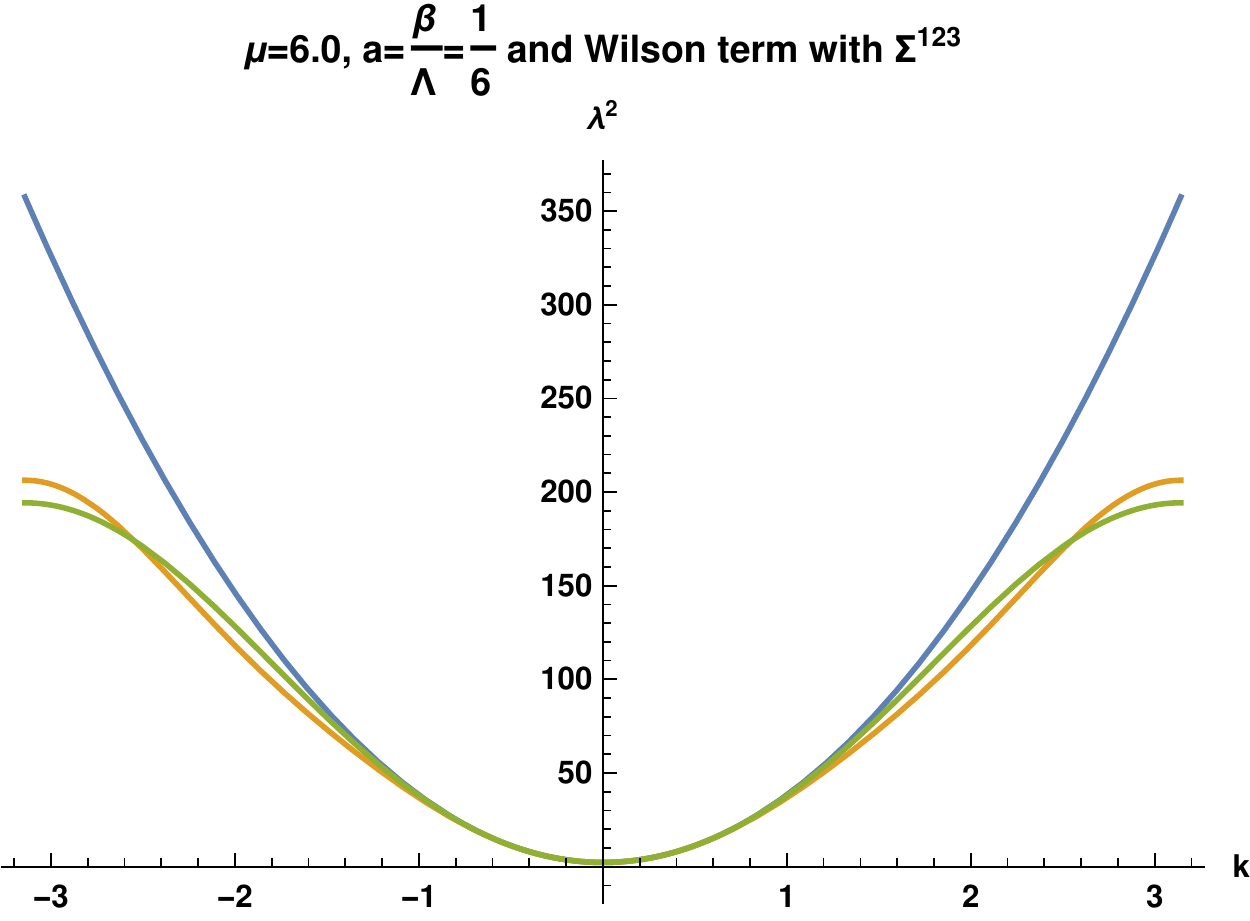}
        \includegraphics[width=2.74in]{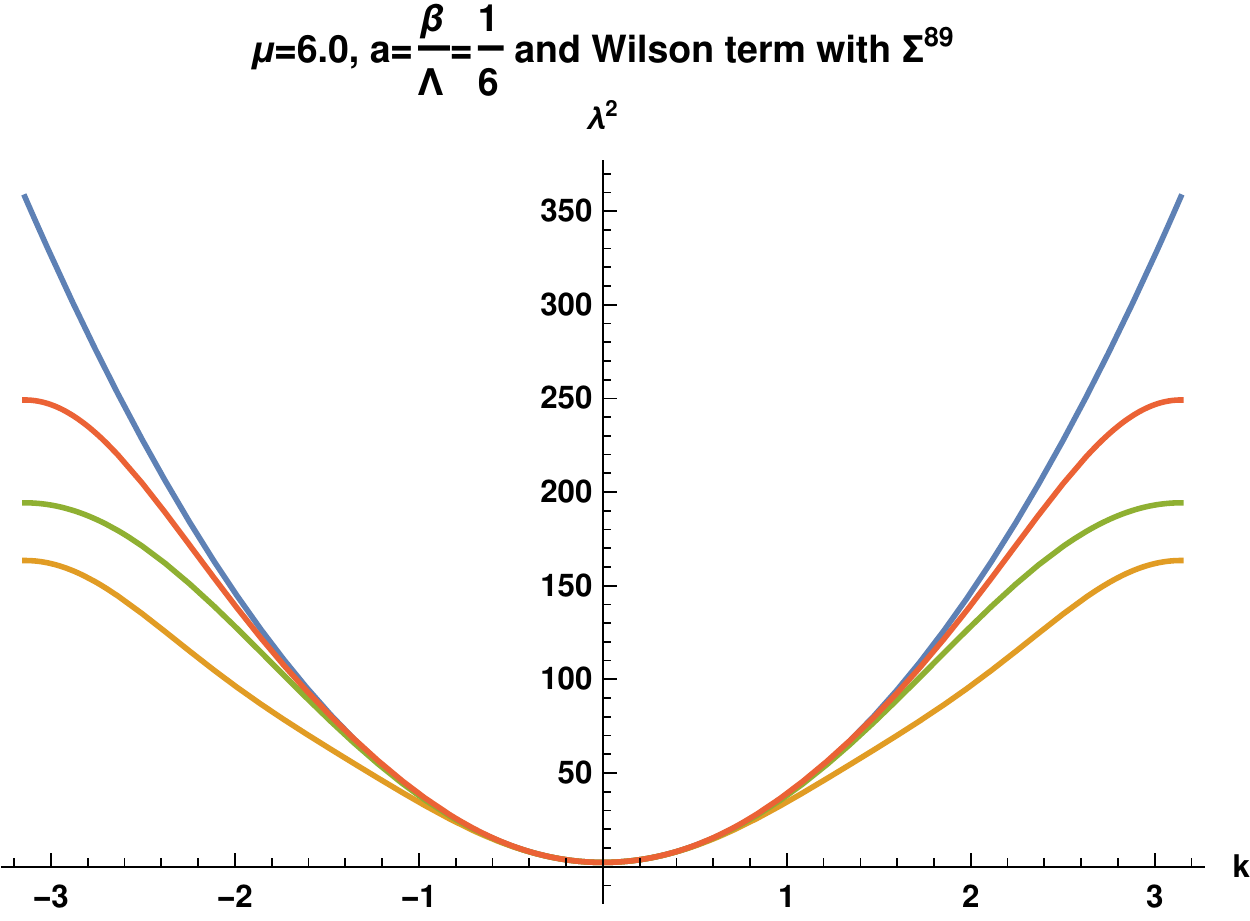}
    \end{tabular}
\caption{\small
  The two figures show the eigenvalues $k^2+\frac{\mu^2}{4}$
  as the blue parabola, and $\Delta_{Bose}+\frac{\mu^2}{4}$ as the
  light green curve. The left figure then shows $D_{Lat}^\dag D_{Lat}$ with
  $\Sigma^{123}$ Wilson term as the orange curve while the right figure
  shows the two distinct eigenvalues of  $D_{Lat}^\dag D_{Lat}$ with $\Sigma^{89}$ Wilson prescription as the red and orange curves. All plots are
  with $\mu=6.0$, $a=\frac{1}{6}$,  $r=-\frac{1}{3}$, $r_{1f}=0$,
  $r_b=\frac{1}{12}$, $r_{2f}=0.1488$, the parameter 
  values used in the simulations and a representative lattice spacing.}
   \label{LatDispersion}
\end{figure}

The bosonic Laplacian is discretised using the lattice version
\begin{equation}
\Delta_{Bose}=\Delta+ r_{b} a^2\Delta^2\, ,
\quad\hbox{where}\quad 
\Delta=\frac{2-e^{a D_\tau}-e^{-a D_\tau}}{a^2}\, . 
\label{BosonicLaplacian}
\end{equation}
In the fermionic action the Dirac operator is discretised as
\begin{equation}
 D_{Lat}=K_a {\bf 1}_{16}- i\frac{\mu}{4} \gamma^{567}+
  \Sigma^{123} K_w\, , \quad\hbox{where}\quad \Sigma^{123}=i\gamma^{123}\, .
  \label{FermionicDiracOperator}
\end{equation}
In (\ref{FermionicDiracOperator})
\begin{equation}
  K_w = r_{1f} a \Delta +r_{2f}a^3\Delta^2
  \end{equation}
is a Wilson term that suppresses fermionic doublers and 
\begin{equation}
 K_a= (1-r)\frac{e^{a D_\tau}-e^{-a D_\tau}}{2a}
    +r\frac{e^{2a D_\tau}-e^{-2a D_\tau}}{4a}\; 
\end{equation}
is a slightly more general discretisation of the derivative
(the standard first order lattice derivative is recovered setting $r=0$).
Here, $K_a$ is an
anti-symmetric lattice operator while $K_w$ is symmetric. The overall
Dirac operator ${\cal C}D_{Lat}$ is an anti-symmetry matrix. The charge conjugation matrix ${\cal C}$ is symmetric as are ${\cal C}\gamma^i$, while ${\cal C}\gamma^{ij}$ and ${\cal C}\gamma^{ijk}$ are the only anti-symmetric elements of the Clifford algebra. An alternative to using $\Sigma^{123}$ would have been to use
$\Sigma^{89}=i\gamma^{89}$ which was the choice used in
\cite{Anagnostopoulos:2007fw,Catterall:2008yz,Hanada,Berkowitz:2016jlq,Catterall:2010gf}
and our earlier code \cite{Filev:2015hia}.
However,  $\Sigma^{89}$, in contrast to $\Sigma^{123}$, does
not anti-commute with the mass term.  Different components of the
spinor then have different dispersion relations and $\mu$-dependent lattice effects which grow with $\mu$. 
We also found that using the $\Sigma^{89}$ prescription and a first order
discretisation (i.e.~where $r, r_b$ and $r_{2f}$ are zero) had
large lattice effects for moderate values of $\mu$.
We discuss details of our code and other technical issues
in \cite{LatticePaper}.  The dispersion relations for
the two options are illustrated in Figure \ref{LatDispersion}, where one
can see two curves for the $\Sigma^{89}$ prescription. As the temperature
is lowered, with fixed $\Lambda$, the lattice spacing becomes larger and
the splitting becomes more extreme.

In practice, the parameter values for our simulations were
$r=-\frac{1}{3}$, $r_{1f}=0$, $r_b=\frac{1}{12}$,
$r_{2f}\approx 0.148$.
We chose $r=-\frac{1}{3}$ as it eliminates the cubic term in
the expansion of $K_a$ and we found that choosing $r_{1f}=0$ minimised
a lattice artifact in the inverse to $D_{Lat}$.  For comparison the
parameters used in \cite{Hanada} were $r=-1$, $r_{1f}=0$,
$r_{2f}=\frac{1}{4}$, $r_b=\frac{3}{4}$ and $\Sigma^{123}$ should be
replaced with $\Sigma^{89}$.

For $\mu=6.0$ and different parameter choices, we performed tests of
our $\Sigma^{123}$ formulation against the $\Sigma^{12}$ option. Fixing
parameters to those of our current study we found results of the
$\Sigma^{123}$ choice with $\Lambda=24$ comparable to the
$\Sigma^{12}$ option with $\Lambda=48$.  

In this paper we concentrate on $\Lambda=24$ and $N=8$; however,
preliminary simulations comparing $\Lambda=24$ and $\Lambda=48$ show
that the lattice effects are small especially for $\mu>4.0$ in the
range of temperatures we studied. More generally, we found no
observable difference in the location of the transitions
between the two lattice sizes.

\section{Observables}
\label{Observables}
\begin{figure}[t] 
  \begin{tabular}{@{}c@{}}
    \includegraphics[width=2.74in]{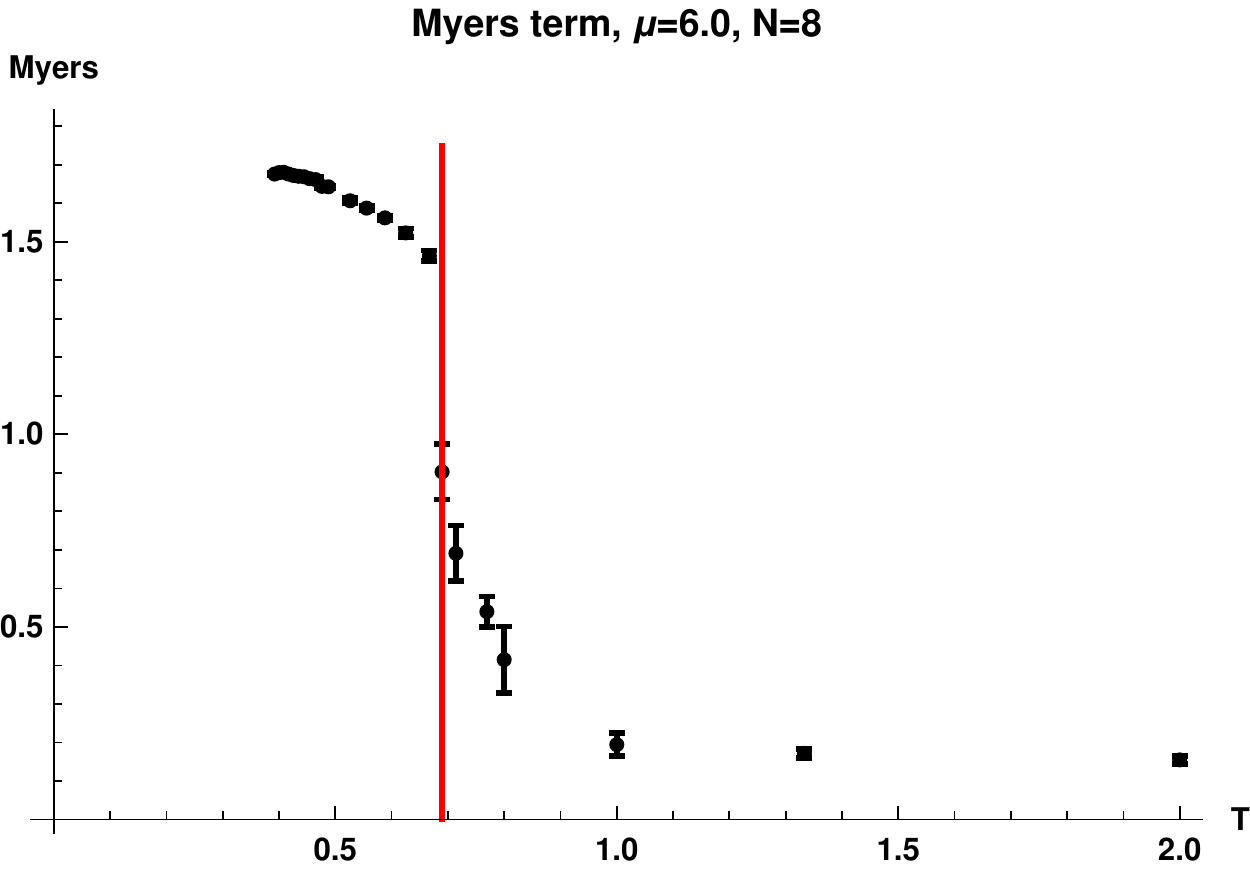}
    \includegraphics[width=2.74in]{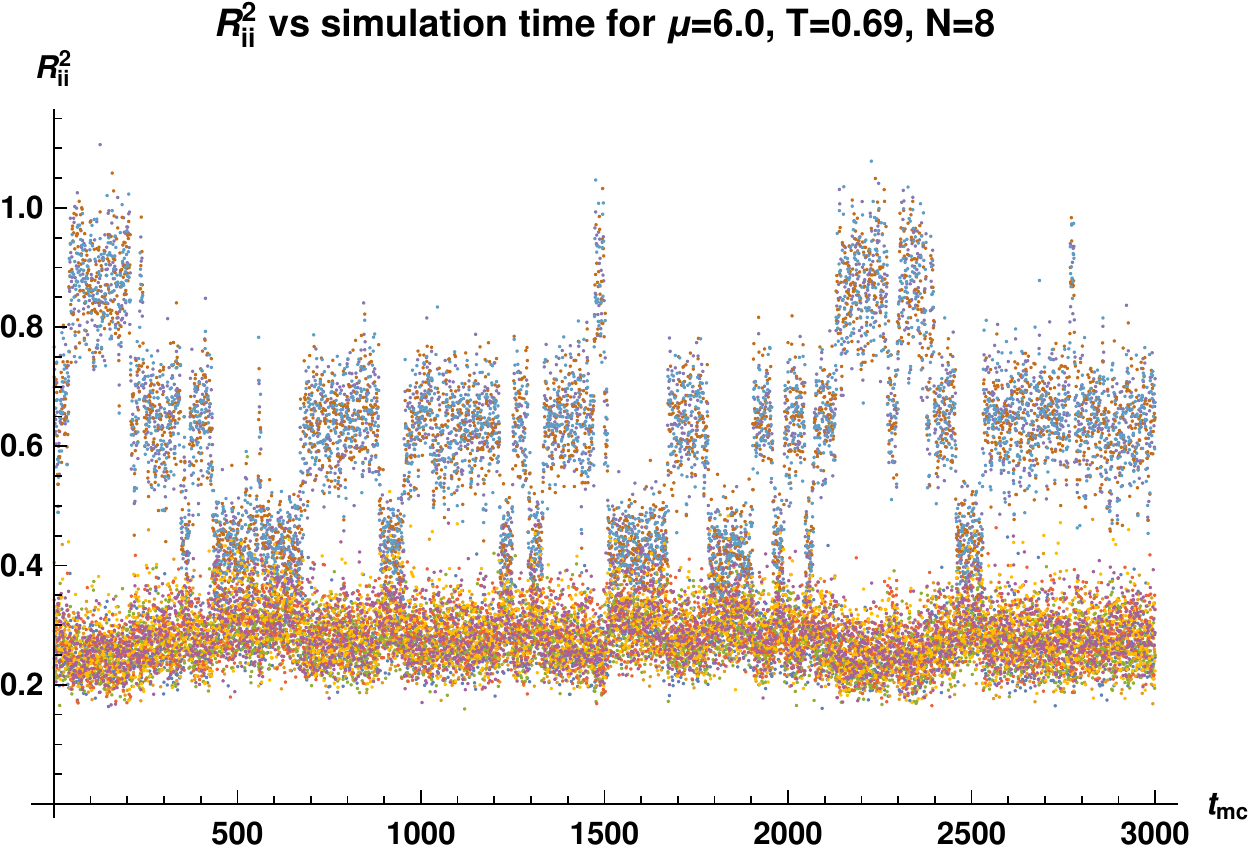}
  \end{tabular}
  \begin{tabular}{@{}c@{}}
    \includegraphics[width=2.74in]{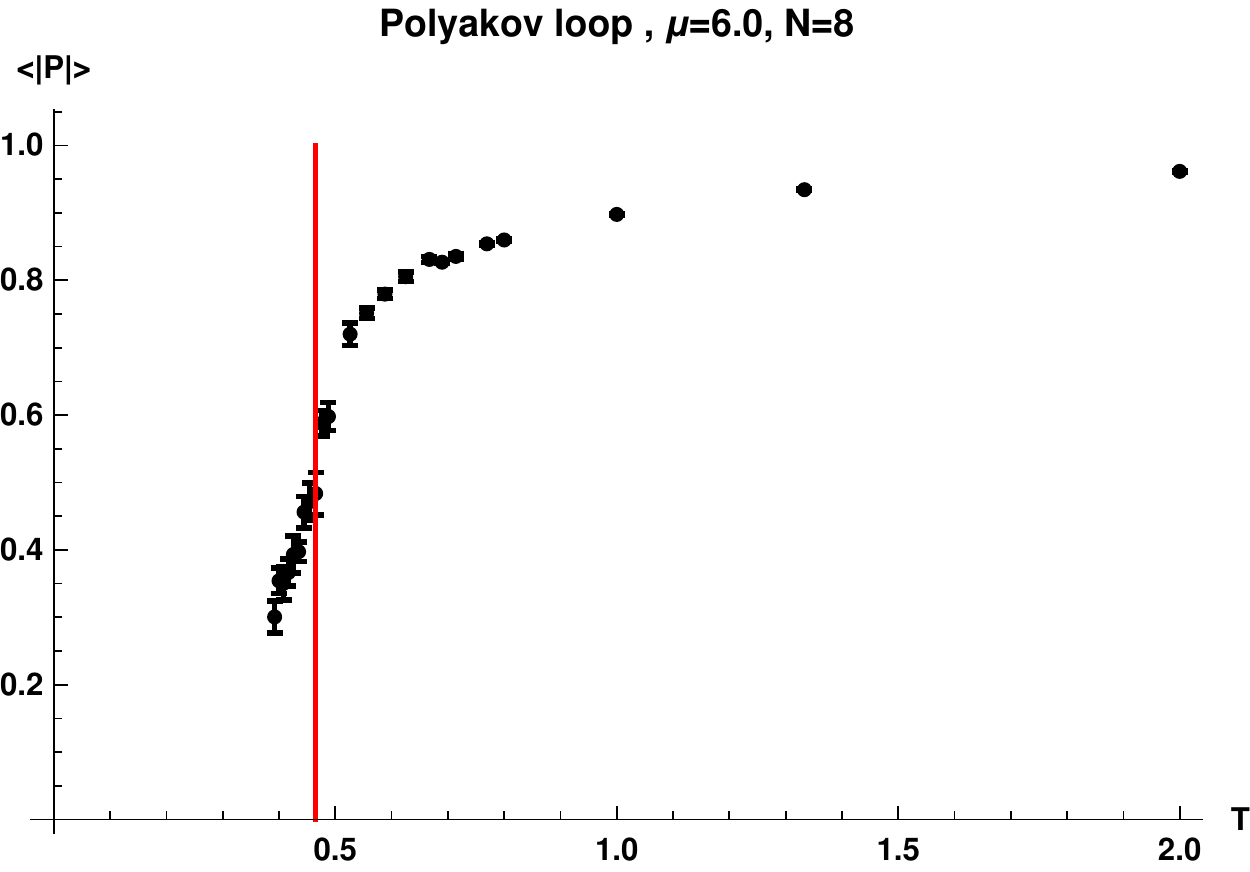}
    \includegraphics[width=2.74in]{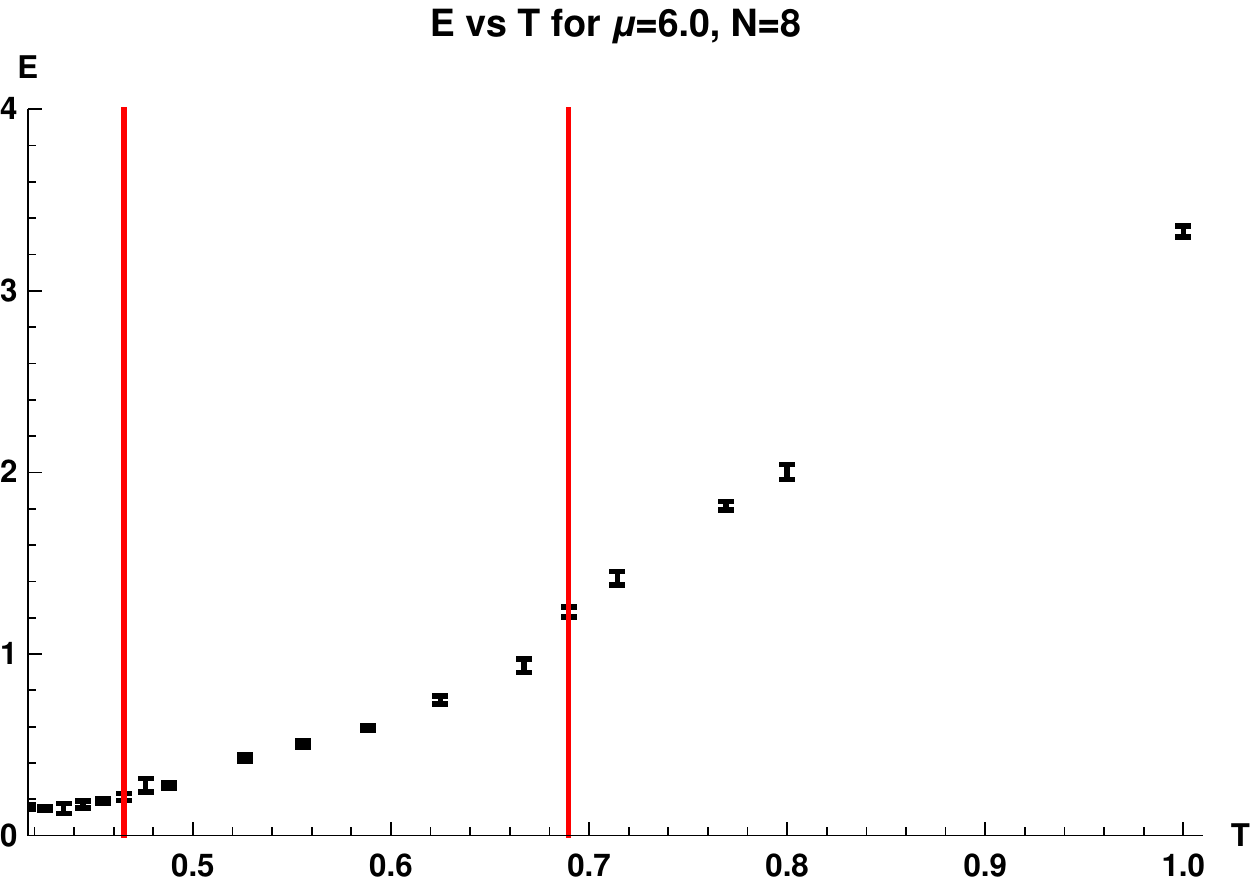}\\
    \includegraphics[width=2.74in]{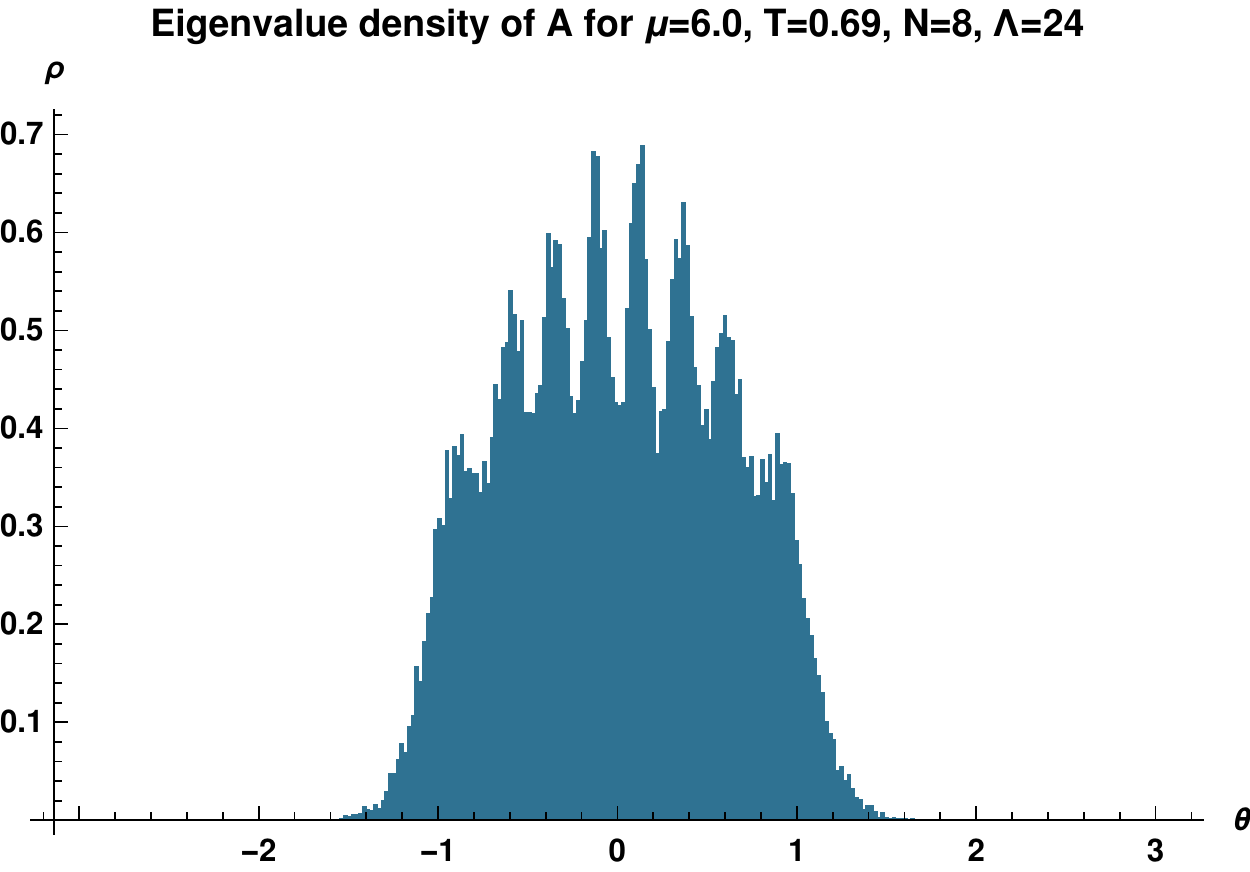}
    \includegraphics[width=2.74in]{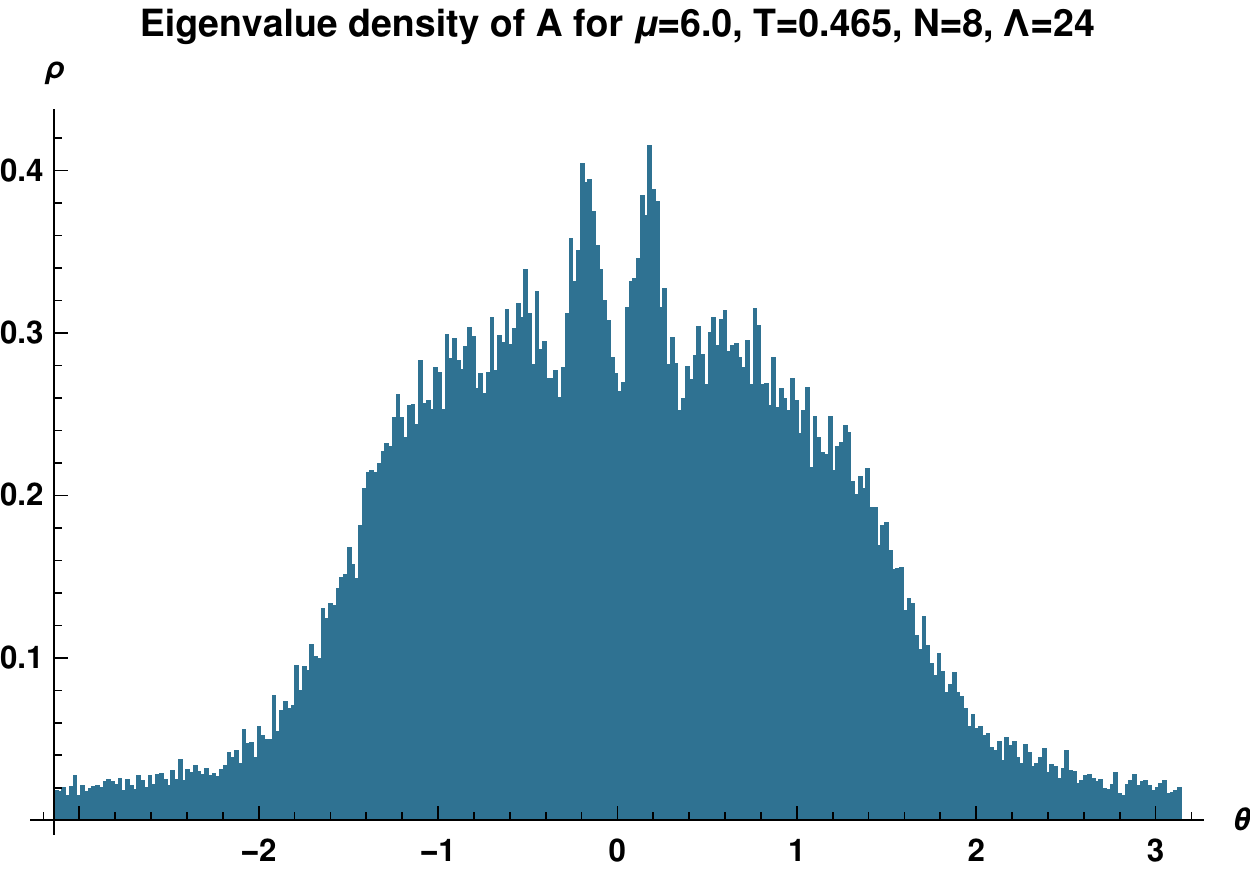}\\
  \end{tabular}
    \caption{\small The figure shows the Myers observable, the
      Polyakov loop and the energy for $\mu=6.0$ and $N=8$.  Two
      transitions occur at distinct temperatures, with the Myers
      transition occurring at $T_{c1}=0.690$, while the Polyakov
      transition, visible in both the Polyakov loop and eigenvalue
      density of the holonomy, occurs at $T_{c2}=0.465$.  The top
      right figure shows $R^2_{ii}$ for each of the nine matrices at
      $T_{c1}=0.690$ marked as the red vertical line in the Myers plot
      and the higher temperature line in the energy plot. The upper points
      correspond to the $SO(3)$ matrices. The Polyakov
      loop shows a transition at $T_{c2}=0.465$, which is marked as the
      second lower temperature red line in the energy plot.  The
      corresponding eigenvalue distributions for the holonomy at both
      transition temperatures are plotted, and one can see that
      Polyakov loop transition involves the eigenvalue distribution
      changing from gapped to ungapped.}
   \label{Myers_Fig_N8μ6p0L24:Energy:Polyakov:R2ii}
\end{figure}
\begin{figure}[t] 
  \begin{tabular}{@{}c@{}}
    \includegraphics[width=2.74in]{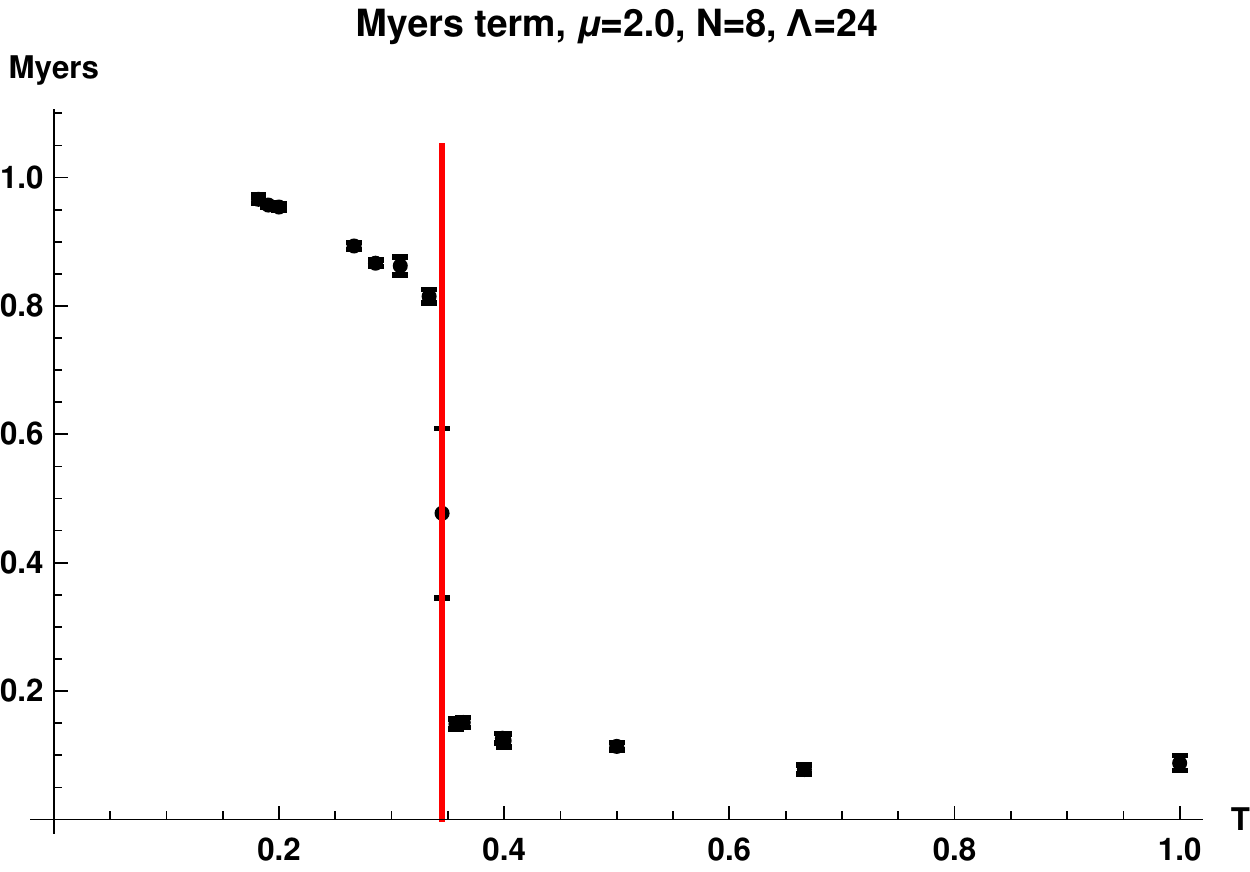}
    \includegraphics[width=2.74in]{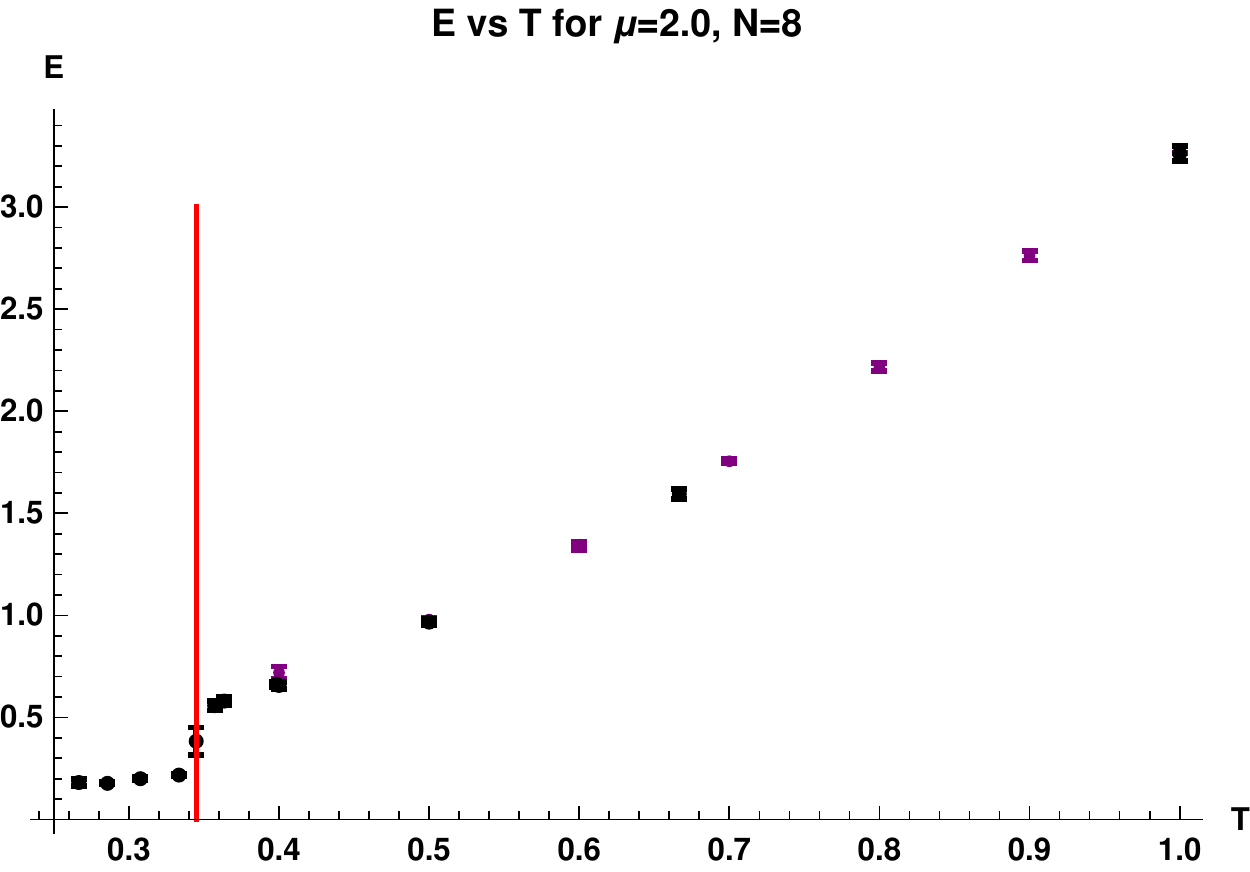}\\
   \includegraphics[width=2.74in]{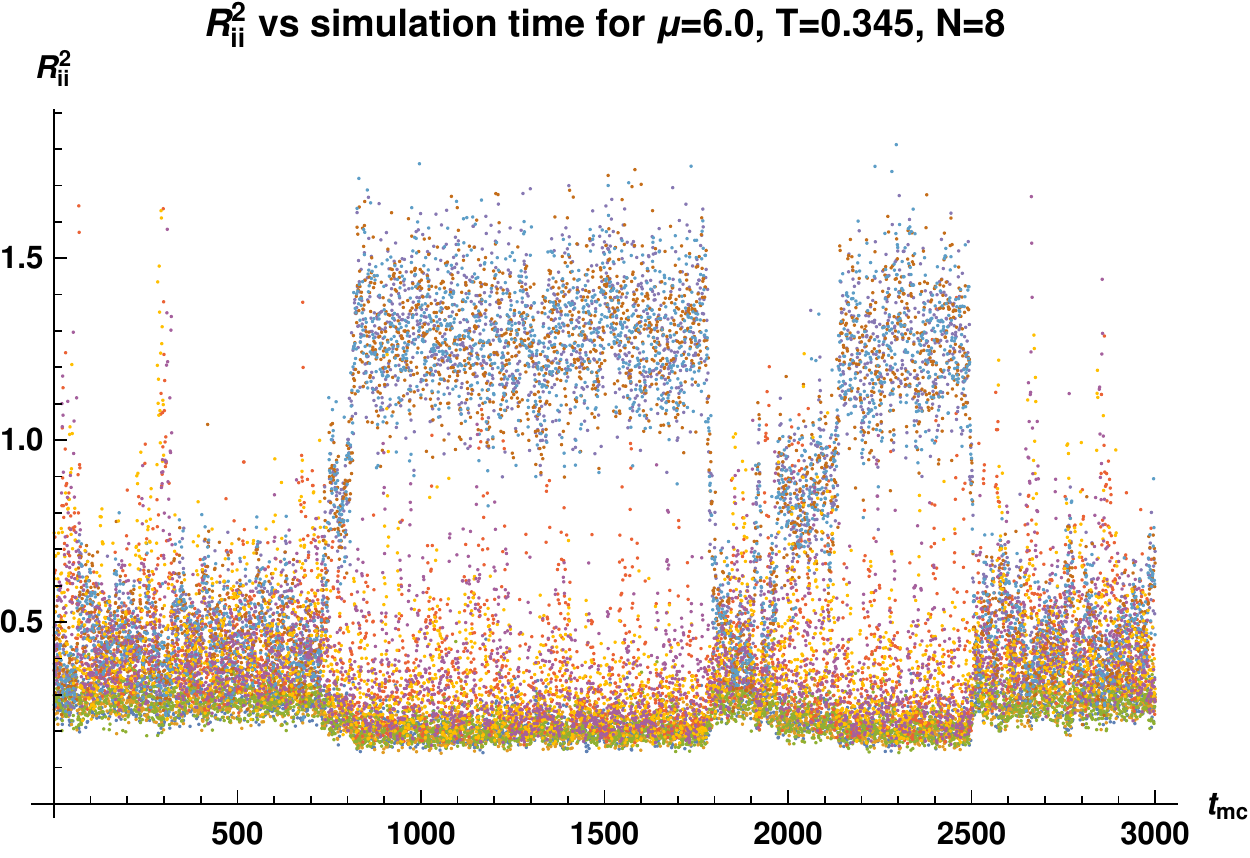} 
    \includegraphics[width=2.74in]{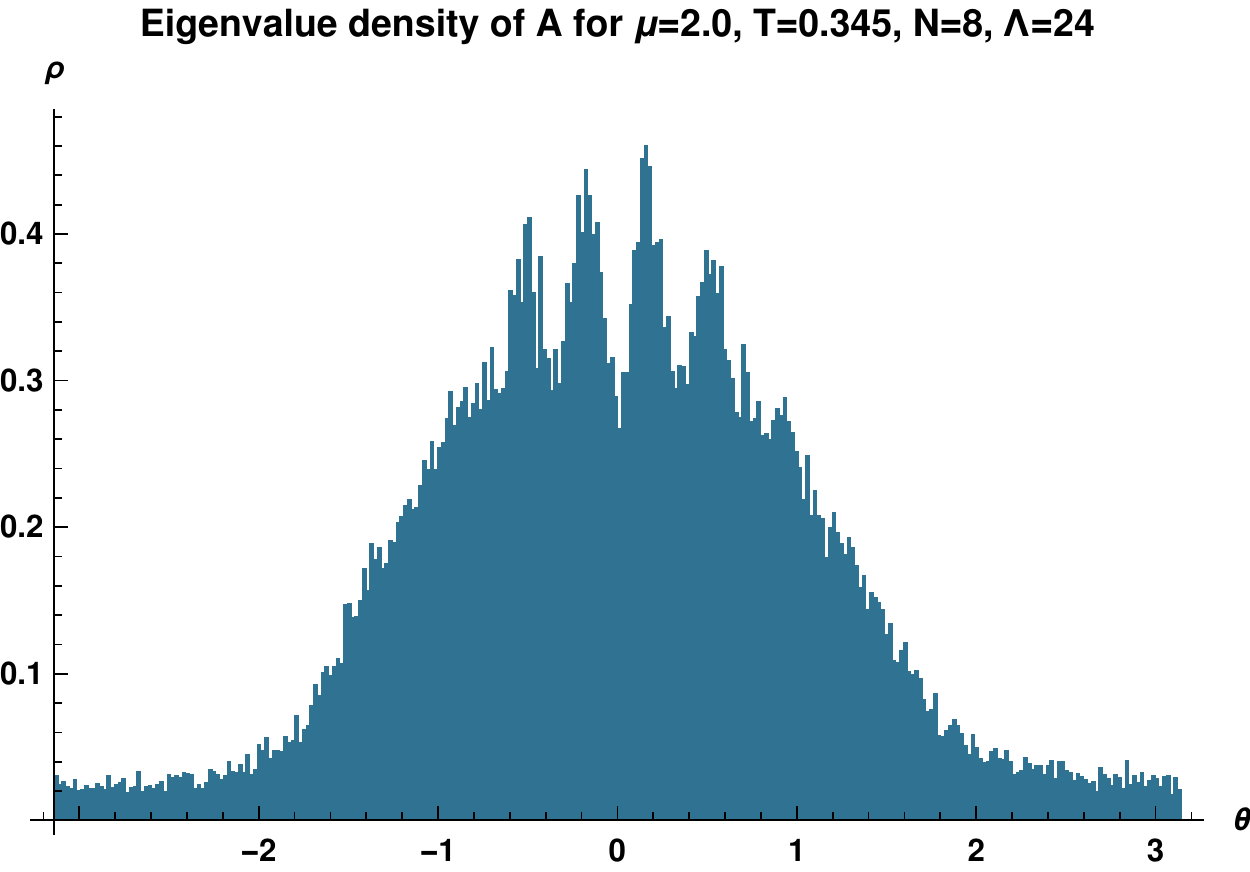} 
\end{tabular}
  \caption{\small The figure shows the Myers observable, the
    energy as a function of temperature, the $R^2_{ii}$ for each $X_i$
    as a function of Monte Carlo sweep and the spectrum of the
    holonomy $\rho$ for $\mu=2.0$, $N=8$, $\Lambda=24$. The red
    vertical line shows our measured value of the transition
    temperature $T_c=0.345$. One can see that the two transitions have
    merged to a single transition for this value of $\mu$. Since the
    mass is relatively small we have also  included, in the energy plot,
    the energy of the BFSS system for $\mu=0$, $N=24$,
    $\Lambda=24$ (the purple points).}
   \label{Energy_Fig_N8μ2p0L24}
\end{figure}
The BMN model is expected to have many phases and a rather complicated
phase structure would be natural at low temperatures due to the
multitude of zero energy BPS states. Our goal in this paper is to map
the high temperature phase boundary to this region. We therefore study
the system as the temperature is lowered, typically beginning our study
at $T=2.0$ for a fixed $\mu$.

In the path integral formulation used in our study one can show,
following the arguments of \cite{Asano:2016xsf}, that the energy is given by
\begin{align}
E=\left\langle\frac{1}{N\beta}\int_0^\beta 
d \tau  \, 
\Tr \Bigg[
  -\frac{3}{4}[X^i,X^j]^2 \right. &-\frac{5i\mu}{6}\varepsilon_{rst}X^rX^sX^t +\left( \frac{\mu}{6}\right)^2(X^m)^2
 \nonumber \\
 & \left. +\left( \frac{\mu}{3}\right)^2(X^r)^2
 -\frac{3}{4} \psi^T\mathcal{C}\gamma^i [X^i , \psi] 
 -\frac{i\mu}{8}\psi^T\mathcal{C}\gamma^{567} \psi
 \Bigg] \right\rangle\, .
 \label{BMNEnergy}
\end{align}
For the BFSS model this reduces to
\begin{equation}
E=\left\langle\frac{1}{N\beta}\int_0^\beta 
d \tau  \, 
\Tr \Bigg[
  -\frac{3}{4}[X^i,X^j]^2 -\frac{3}{4} \psi^T\mathcal{C}\gamma^i [X^i , \psi]  \Bigg] \right\rangle,
 \label{BFSSEnergy}
\end{equation}
and (\ref{BFSSEnergy}) together with the Ward identity, which reads
\begin{equation}
2\langle S_{bos}\rangle + N\int_0^\beta 
d \tau  \, 
\Tr \Bigg[
  -\frac{1}{2}[X^i,X^j]^2 -\frac{1}{2} \psi^T\mathcal{C}\gamma^i [X^i , \psi]  \Bigg]=9(N^2-1)\Lambda\, ,
\label{WardIdentity}
\end{equation}
was typically used to express the energy without fermionic terms as
\begin{equation}
E=\frac{3}{N^2\beta}\left(\frac{9}{2}\Lambda (N^2-1)-\langle S_{bos}\rangle\right).
\label{BFSSEnergyLat}
\end{equation}
While this expression provides coding convenience it is not necessary
as one can measure these fermionic observables using pseudo
fermions. Also, note that (\ref{BFSSEnergyLat}) involves the
difference of large numbers and is limited by the precision of the
rational approximation to the fermionic contributions. Using this
expression to determine the energy in a precision calculation can
become a problem especially when approaching the continuum limit, a
limit involving sending $\Lambda$ to infinity.

\begin{figure}[t] 
  \begin{tabular}{@{}c@{}}
        \includegraphics[width=2.74in]{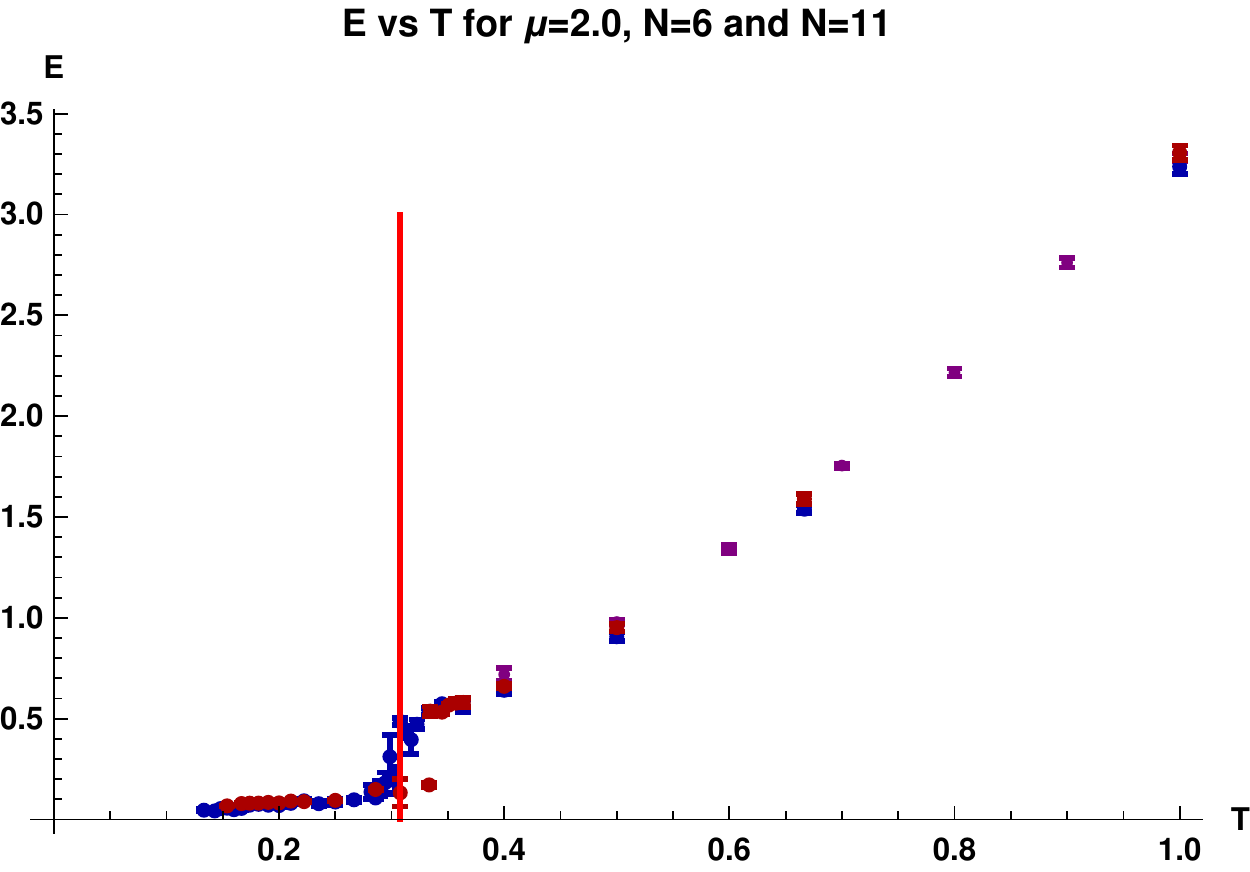}
    \includegraphics[width=2.74in]{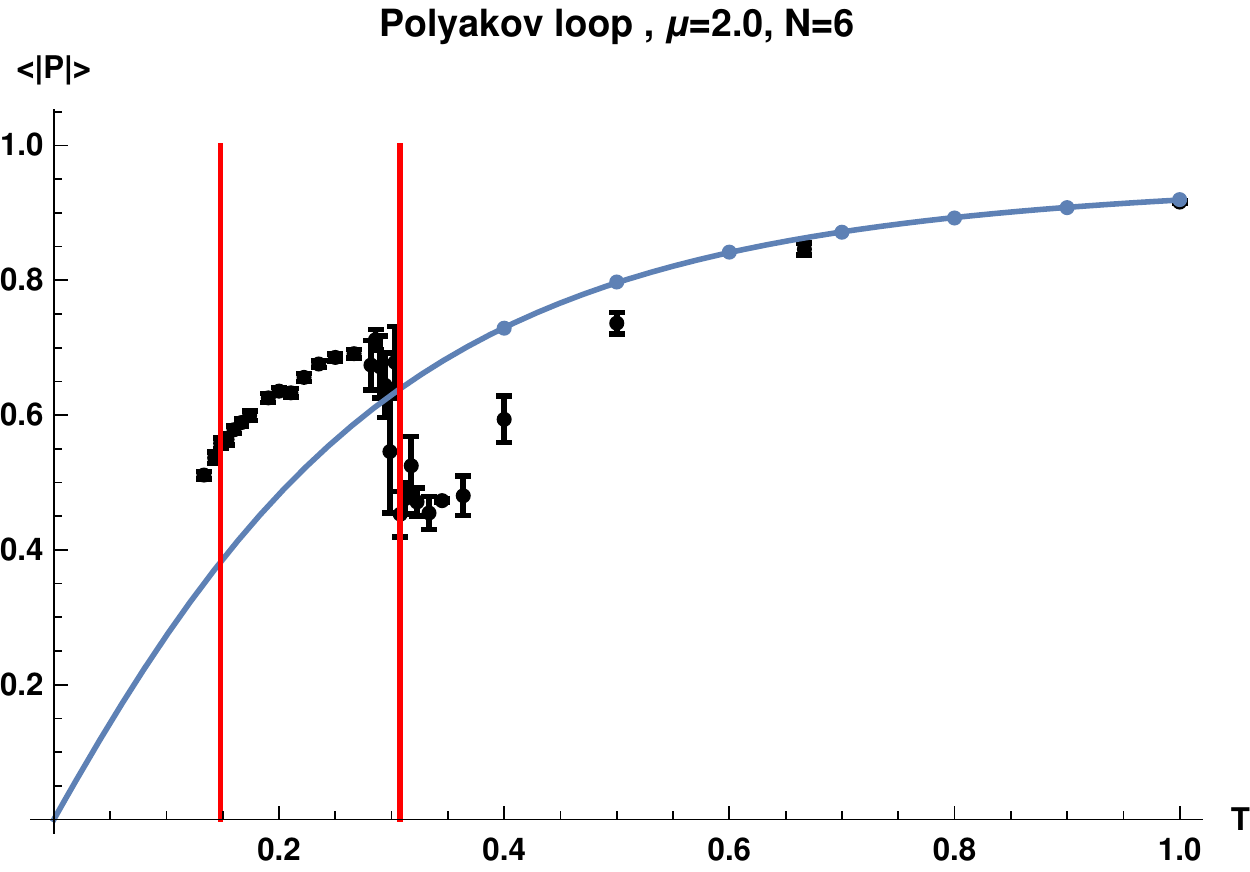} 
\end{tabular}
  \caption{\small The left figure shows superimposed energy curves for
    $N=6$ and $N=11$. The right figure shows $\langle|P|\rangle$
    for $N=6$, $\mu=2.0$ and $\Lambda=24$. The $N=6$ transition is
    taken to be at
    $T_c=0.31\pm 0.02$. Both plots include the BFSS energy
    (i.e. $\mu=0$) and $\langle|P|\rangle$ values for $N=24$,
    $\Lambda=24$. This data was used in the $1/N$
    extrapolation in Figure \ref{mu-T_Phase_diagram}.
    }
   \label{EnergyAndPol_N6_μ2p0L24}
\end{figure}
\begin{figure}[t] 
   \begin{tabular}{@{}c@{}}
   \includegraphics[width=2.74in]{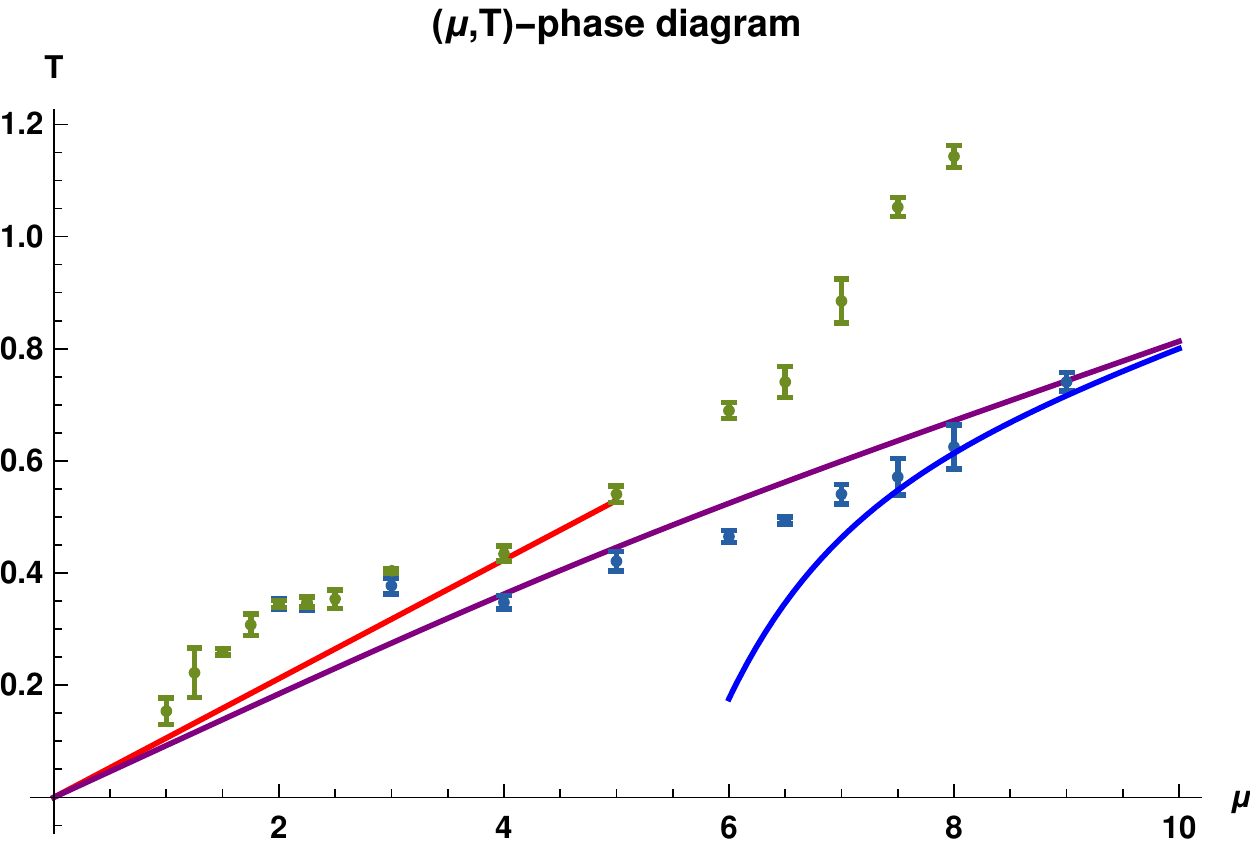}
   \includegraphics[width=2.74in]{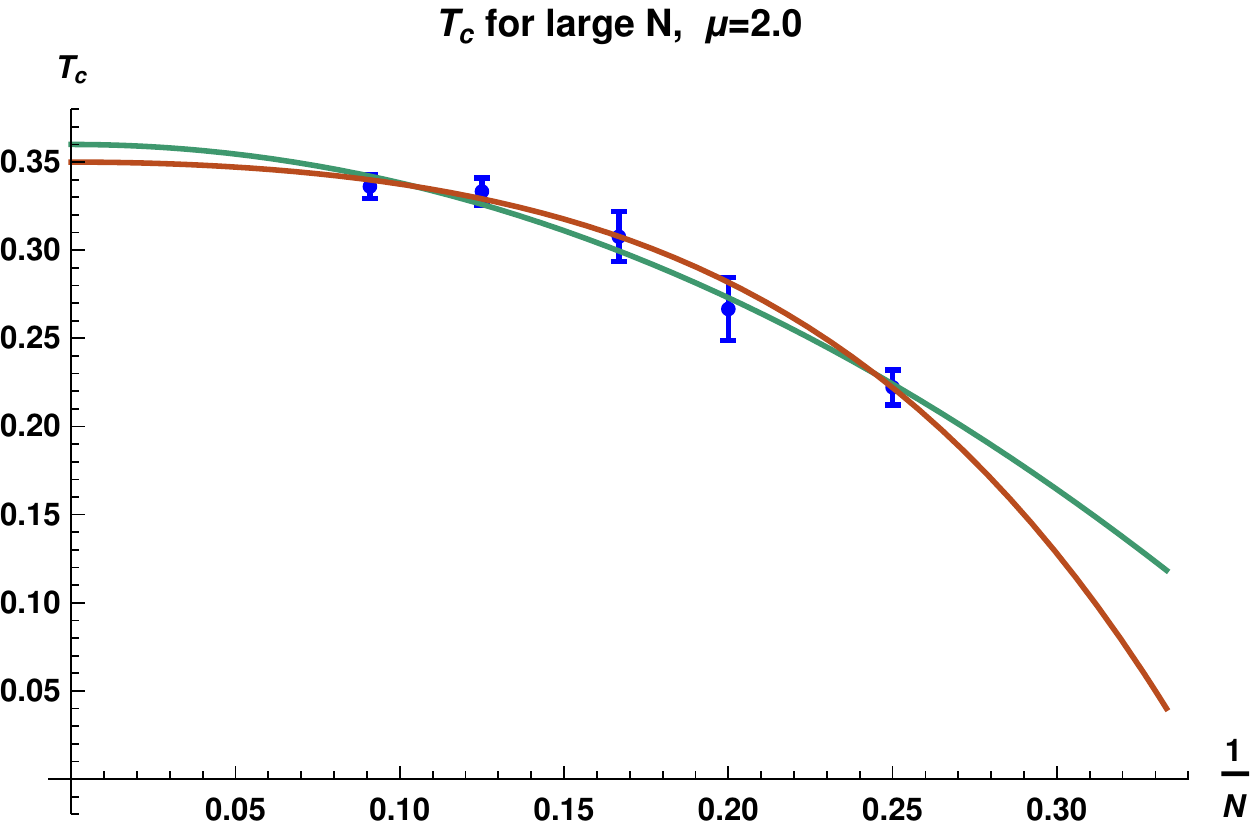}
   \end{tabular}
   \caption{\small The left figure shows data for the two observed
     phase transitions. The green points are the observed values of
     the Myers transition and the dark blue points are the Polyakov
     transition, both were measured for $N=8$ and $\Lambda=24$. The
     blue curve is the large $\mu$ expansion of the critical
     temperature to 3 loops. The red line is the gauge gravity
     prediction and the purple curve is the Pad\'e approximant obtained
     from the large mass expansion.\hfill\break
     The figure on the right shows our extrapolation of the critical
     temperature of the Myers transition from the observed values for
     $N=4,5,6,8$ and $N=11$ for $\mu=2.0$ and $\Lambda=24$.  We use a
     quadratic extrapolation (green), given by $0.36-2.17x^2$ and
     quartic (red) given by
     $0.35 - 1.09x^2-15.26x^4$, which gives $T_c(\infty)=0.35\pm0.01$.  }
   \label{mu-T_Phase_diagram}
\end{figure}

In practice we find that the direct observable (\ref{BFSSEnergy}) behaves
better numerically.  For the BMN model
we implement (\ref{BMNEnergy}) directly using pseudo fermions
following the strategy discussed in
section 4.2 of \cite{Filev:2015hia} and used in \cite{Asano:2016kxo}.
We also, for completeness, observe the BMN Ward identity
analogue of (\ref{WardIdentity}).

The observables we follow are then: $E$ as defined in (\ref{BMNEnergy}),
\begin{align}
 {\rm Myers}
 &=\left\langle \frac{i}{3N\beta}\int_0^\beta d\tau \epsilon_{rst}\Tr(X^rX^sX^t) \right\rangle\, ,
 \label{Myers}\\
 \left\langle \vert P\vert\right\rangle
 &= \left\langle \frac{1}{N} \vert\Tr \left( \exp \left[ i \beta A \right] \right) \vert\right\rangle
 \, ,
 \label{PolyakovLoop} \\
 \quad\qquad R^2_{ii} 
 &= \left\langle \frac{1}{N\beta}\int_0^\beta d\tau \Tr(X^{i}X^{i})\right\rangle 
 \quad\hbox{(no sum on $i$).}
 \label{R2ii}
\end{align}

The energy and specific heat of the system, which due to supersymmetry,
should both decrease to zero as $T$ approaches zero, are given by
\begin{equation}
  N^2 E=\langle H\rangle=-\partial_{\beta}\ln Z
   \quad\hbox{ and}\quad 
  C=\langle(H-\langle H\rangle)^2\rangle=\partial^2_{\beta}\ln Z ~\geq~0.
\end{equation}
We see, furthermore, that $E$ must be a monotonic
function of $T$. This proves especially useful in the simulations as
tracking the energy as a function of $T$ was a crucial clue in
identifying when the system transitioned to a new level\footnote{If
  the system was not cooled in sufficiently small temperature steps,
  $E(T)$ was sometimes found to be slightly larger at the lower
  temperature. This indicated that the lower temperature state was not
  the thermodynamically preferred one.}.  With this strategy and
careful simulation as the transition was approached we identified the
phase diagram as shown in Figure~\ref{mu-T_Phase_diagram}.

\section{Results and Phase Diagram}
\label{ResultsAndPhaseDiagram}
For large $\mu$, e.g.~$\mu=18.0$, we find that the transition is well
tracked by the 3-loop perturbative result (\ref{LargemuTc}); however,
deviations arise as $\mu$ is reduced.  We concentrate our efforts on
$\mu\leq 9.0$. We find that for $\mu\leq 6.5$ the system undergoes a
phase transition from a small Myers observable to a large one and
there is no longer an approximate $SO(9)$ symmetry.  This is clear
when observing $R^2_{ii}$, which at temperatures close to the
transition has large fluctuations in the $SO(3)$ components (see
Figure~\ref{Myers_Fig_N8μ6p0L24:Energy:Polyakov:R2ii}). A reasonable
order parameter for this transition is the Myers observable
(\ref{Myers}), also shown in
Figure~\ref{Myers_Fig_N8μ6p0L24:Energy:Polyakov:R2ii} for $\mu=6.0$ as
a function of temperature.

Figure~\ref{Myers_Fig_N8μ6p0L24:Energy:Polyakov:R2ii} shows there are
two separate transitions with the Myers transition occurring at the
higher temperature. The figure also shows $R^2_{ii}$ for each of the
nine matrices during a run corresponding to the transition value
$T_{c1}=0.690$ marked as the red vertical line in the Myers plot and
the higher temperature line in the energy plot. One can see that the
$SO(3)$ matrices fluctuate between different fuzzy sphere vacua, while
there are only small variations in the $SO(6)$ matrices, which are
concentrated at smaller values.

The deconfinement transition is clear in
the plot of the Polyakov loop and is measured to occur at
$T_{c2}=0.465$. It is the second red line marked in the energy
plot. The corresponding eigenvalue distributions of the holonomy for
the two transition temperatures are plotted below and one can see that
the confinement-deconfinement transition involves the eigenvalue
distribution changing from gapped to ungapped.
One can also see evidence for both transitions in the plot
of $E$, though the Polyakov
transition is less clear due to the relatively small jump in
energy across this transition. However, as the temperature is lowered
the transition in the energy becomes more pronounced and the two
transitions approach, eventually effectively merging in the vicinity of
$\mu=3.0$.  Figure~\ref{Energy_Fig_N8μ2p0L24} shows our observables
for $\mu=2.0$ with $N=8$ and $\Lambda=24$. Only one transition was
observed and this transition is clear in the energy plot.

Once the system is cooled to a temperature below $T_{c1}$ the $SO(3)$
matrices settle into the thermodynamically favoured fuzzy sphere configuration,
and the $SO(6)$ ones fluctuate around smaller
values.
Though larger fuzzy sphere vacua exist, further transitions between
fuzzy sphere configurations were not observed as the system was cooled
further.

We present our overall measured phase diagram in
Figure~\ref{mu-T_Phase_diagram}. The figure is restricted to $N=8$
with $\Lambda=24$ and was determined from the Myers observable,
Polyakov loop and
energy observables. The simulations were performed beginning at
$T=2.0$; then $T$ was slowly decreased for fixed $\mu$ using the
thermalised input obtained at the higher temperature as an initial
configuration for the lower $T$.  The strategy of slowly decreasing
the temperature for fixed $\mu$ proved crucial in determining the
transition, as in many cases hysteresis made it difficult for the
simulation to find the correct phase at the lower temperature. Which
phase was correct was determined by carefully observing that the
energy remained a monotonic function of $T$. The
simulations indicate that the observed transitions are all first
order, with a significant latent heat for $\mu=2.0$.

In the second graph in Figure~\ref{mu-T_Phase_diagram} we focus on $\mu=2.0$ and
extrapolate the critical temperature to large $N$ estimating that
$T_c(\infty)=0.35\pm0.01$, where the measured value for $N=8$ was
$T_c(8)=0.345\pm0.006$. From this example and preliminary results for
other $\mu$ we expect that the large $N$ curve is shifted upwards slightly
towards higher temperatures and we infer that the transition curve as
measured in Figure~\ref{mu-T_Phase_diagram} is a lower estimate for
the large $N$ coexistence curve.
Figure \ref{EnergyAndPol_N6_μ2p0L24} shows the energy and Polyakov
loop for $N=6$, with $\mu=2.0$. We see that the transition occurs at
$T_c(6)=0.31\pm0.02$, which is lower than $T_c(8)$; both are used in
the $\frac{1}{N}$ extrapolation of Figure \ref{mu-T_Phase_diagram}.
Figure \ref{EnergyAndPol_N6_μ2p0L24} also shows the Polyakov loop and
we see it makes a sharp transition to a larger value. From this we
infer that the fuzzy sphere configurations have typically larger
Polyakov loop at a given temperature than the approximately $SO(9)$ symmetric phase. We
further find that in this phase the holonomy becomes ungapped at
$T\sim0.15\pm0.02$.
\section{Conclusions}
\label{Conclusions}

We have mapped an initial phase diagram for the BMN matrix model in
the $(\mu,T)$ plane. We have concentrated on relatively small matrix
sizes, with most of our data for $N=8$.  The resulting diagram is
presented in Figure~\ref{mu-T_Phase_diagram}. 

We found that for small enough $N$ and
large enough temperature ergodicity is not a problem. However,
for large $N$ ergodicity is lost in the simulations, we therefore
restricted our study to small $N$.
For $N=6$ we tracked the system to relatively low
temperatures (down to $T\sim 0.1$ with $\mu=2.0$). A little below the
transition temperature it was apparent that maintaining ergodicity
was beginning to be a problem as transitions between levels required rather
lengthy simulations. For $N=8$ and $N=11$ there
were difficulties with ergodicity, however, cooling
the system in sufficiently small temperature intervals allowed us to
access the transition. We suspect that going beyond
$N=16$ would require new techniques to overcome these difficulties.

We have observed two transitions for larger values of $\mu$, which
appear to merge for $\mu\sim 3.0$. The higher temperature transition
curve, which shows the Myers transition curve,
is the transition from the trivial vacuum to a non-trivial fuzzy
sphere vacuum. In this
transition the ground state consists of the $SO(3)$ matrices blowing
up into fuzzy spheres and the transition has the characteristics of
that discussed in
\cite{DelgadilloBlando:2007vx,DelgadilloBlando:2008vi,OConnor:2013gfj}.
The Myers transition became very difficult to observe for $\mu>6.5$ as
it occurred in a very narrow temperature interval.  We have not
successfully tracked it to very large $\mu$ and we suspect that, at
least in the large $\mu$ region of the phase diagram, it is a
finite matrix effect.

The second transition is observed in the Polyakov loop and is
associated with the eigenvalue density of the gauge field, or
holonomy, transitioning from gapped to ungapped (from covering a small
interval of the unit circle to covering the entire circle).  For
$\mu\sim 6.0$ this occurs in a narrow temperature interval. However,
as $\mu$ is lowered the Polyakov loop is not a monotonic function of
temperature; it jumps upward at the Myers transition, where the system
transitions to larger fuzzy sphere configurations. Each BPS ground
state has its own typical value for the holonomy. In the Hamiltonian
picture the holonomy implements the Gauss law constraint and is
therefore sensitive to the density of accessible $SU(N)$ non-singlet
states\footnote{For a recent discussion of the r\^ole of the gauge
  filed see \cite{Maldacena:2018vsr} and
  \cite{Berkowitz:2018qhn}.}. It plays an essential r\^ole when there
are many non-singlets around the relevant energy.  One would expect
that different fuzzy sphere configurations have different densities of
non-singlets, which would account for the behaviour of the holonomy,
as it is effectively probing these.

We found that as $\mu$ was decreased the two transitions merge at
about $\mu\sim 3.0$.  This single transition curve then approaches the
Pad\'e prediction (\ref{PadeResummedTc}) and the gauge/gravity
\cite{Costa:2014wya} prediction (\ref{gaugegravity}). We have not 
tracked the transition to $\mu$ less than $1.0$.\footnote{Tracking the
  transition to very small $\mu$ would also require larger lattices
  and larger matrix size, $N$.}

We find strong evidence that the Polyakov loop transition
coincides with the Myers transition for $\mu<3.0$. Also, our results
are in qualitative agreement with the gravity predictions of
\cite{Costa:2014wya}. However, we expect that better agreement over a larger
range of $\mu$ would be obtained should a black hole
solution dual to a general vacuum be available.

Another remarkable observation is that, at strong coupling, $\mu=2.0$,
the Myers term seems to have a finite, non-zero value in the large-$N$
extrapolation.  If one considers a typical vacuum
as a representation with $m$ copies of the $k$
dimensional representation, it yields at low
temperatures
$\mathrm{Myers}\sim \frac{\mu^3}{3^4N}\Tr(J^rJ^r)\sim \mu^3(k^2-1)$.
Such a configuration provides an example where this observable
is non-zero and does not diverge, when $k$ is finite.  Thus the easiest way
to realise configurations that have finite Myers term at large $N$ is
to have many copies of representations that fluctuate around a typical
one of relatively small dimension.  This should correspond to a state
of five-branes \cite{Maldacena:2002rb}, and in turn suggests that the
Myers tranitions for $\mu<3$ can be regarded as a transition to a
five-brane phase.

Our current study is clearly only the beginning. However, it indicates
that an effort to construct such a black hole configuration
in the dual gravitational theory, even a numerical one, would be very
useful.

In the near future we plan to add D4-brane probes to the BMN
model \cite{Kim:2002cr} in analogy with our studies \cite{Asano:2016kxo,Filev:2015cmz}.
This should give an alternative, more detailed probe, of the dual geometry.

\section*{Acknowledgment}
Part of the simulations were performed within the ICHEC
``Discovery'' projects ``dsphy009c'' and ``dsphy011c'', the DIAS cluster and DIAS ICHEC nodes under ``dias01''.
The support from Action MP1405 QSPACE of the COST foundation
is gratefully acknowledged. Y.A.~thanks M.~Hanada, G.~Ishiki and J.~Nishimura for valuable discussions.
Y.A.~is supported in part by the JSPS Research Fellowship for Young Scientists.
S.K.~was supported by Irish research council funding. 
The work of V.~F. and D.~O. was supported in part by the Bulgarian NSF grant DN08/3.

\end{document}